\documentclass[aps,preprint,pra]{revtex4}
\usepackage{graphicx}

\begin{document}

\title{PR-box correlations have no classical limit}\thanks{{\it Quantum Theory:  A Two-Time Success Story} (Yakir Aharonov Festschrift), eds. D. C. Struppa and J. M. Tollaksen (New York:  Springer), 2013, pp. 205-211. For a video of this talk at the Aharonov-80 Conference in 2012 at Chapman University, see quantum.chapman.edu/talk-10.}

\author{Daniel Rohrlich}
\affiliation{Department of Physics, Ben Gurion University of the Negev, Beersheba
84105 Israel}

\date{\today}

\begin{abstract}
One of Yakir Aharonov's endlessly captivating physics ideas is the conjecture that two axioms, namely relativistic causality (``no superluminal signalling") and nonlocality, so nearly contradict each other that a unique theory---quantum mechanics---reconciles them.  But superquantum (or ``PR-box") correlations imply that quantum mechanics is not the most nonlocal theory (in the sense of nonlocal correlations) consistent with relativistic causality.  Let us  consider supplementing these two axioms with a minimal third axiom:  there exists a classical limit in which macroscopic observables commute.  That is, just as quantum mechanics has a classical limit, so must any generalization of quantum mechanics.  In this classical limit, PR-box correlations $violate$ relativistic causality.  Generalized to all stronger-than-quantum bipartite correlations, this result is a derivation of Tsirelson's bound without assuming quantum mechanics.
\end{abstract}

\pacs{03.65.Ta, 03.30.+p, 03.65.Ud, 03.67.Hk}

\maketitle

I first met Yakir Aharonov when I was a post-doc at Tel Aviv University, more than two decades ago.  I discovered that he could answer questions about quantum mechanics that no one had answered to my satisfaction, not even Nobel Prize-winning physicists.  I wanted to understand what he understood about quantum mechanics, and gladly accepted his offer to write a book together \cite{QP}.  Around that time, he told me an idea that has fascinated me ever since.  What follows is the story of that idea.

Of course quantum mechanics is baffling, he said.  Look at its axioms:  ``Physical states are normalized vectors in Hilbert space"; ``Measurable physical quantities correspond to self-adjoint operators"; the Born rule; etc.  Are these statements about the physical world, or statements about applied mathematics?  By contrast, special relativity has an exemplary logical structure:  two axioms, each with a clear physical meaning---a fundamental physical constant (the speed of light) and a fundamental invariance (the equivalence of inertial reference frames)---are so nearly incompatible that a unique kinematics reconciles them.  Suppose we tried to derive special relativity from the wrong axioms, e.g. ``Fast objects contract in the direction of their motion" and ``Moving clocks slow down."  How could we understand the theory?  If the analogy seems far-fetched, note that as late as 1909, Henri Poincar\'e gave a lecture entitled ``La m\'echanique nouvelle" (``The new mechanics").  The lecture, after mentioning the two axioms of special relativity, goes on to mention the FitzGerald contraction:  ``One needs to make still a third hypothesis, much more surprising, much more difficult to accept, one which is of much hindrance to what we are currently used to. A body in translational motion suffers a deformation in the direction in which it is displaced \cite{hp}."  The phrase ``of much hindrance"---wouldn't it apply to any of the axioms of quantum mechanics?!  The point here is that for Poincar\'e, the FitzGerald contraction was an axiom and therefore hard to accept; for us, the FitzGerald contraction is a logical consequence of clear physical axioms and therefore not so hard to accept.

But, Aharonov continued, quantum mechanics, too, reconciles two nearly incompatible physical statements.  On the one hand, quantum mechanics has to obey {\it relativistic causality}, the constraint that no signal can travel faster than light.  On the other hand, quantum mechanics is nonlocal, in at least two different ways.  There are the Aharonov-Bohm \cite{AB} and related effects, in which the motions of particles depend on nonlocal relative phases; and there are nonlocal quantum correlations that violate a Bell inequality
\cite{bell, chsh}.  Quantum nonlocality is action at a distance:  a cause $here$ has an immediate effect $there$.  How can action at a distance be compatible with relativistic causality?  The only way to reconcile them is via {\it uncertainty}.  If the effect $there$ of a cause $here$ is uncertain, it may not lead to superluminal signalling.  Thus we can obtain uncertainty as a logical consequence of axioms of relativistic causality and nonlocality.  It is hard to accept uncertainty as an axiom (the Born rule); it leaves us asking, ``Why does God play dice?"  But it is not so hard to accept uncertainty as a logical consequence of relativistic causality and nonlocality; we can say, ``God plays dice because it is the only way for nonlocality and relativistic causality to coexist."

Is this derivation of quantum uncertainty only qualitative, or is it also quantitative?  To answer this question, we must specify what we mean by quantum nonlocality.  Aharonov defined ``modular" quantum variables that are nonlocal in space or time.  The nonlocality of modular variables arises from nonlocal relative phases.  He showed that there is always just enough uncertainty in measurements of modular variables to prevent their use for noncausal signalling \cite{nleq}.  Independently, Shimony noted that quantum mechanics is remarkable in reconciling relativistic causality with the nonlocality implicit in nonlocal quantum correlations.  He gave this coexistence the apt name ``passion at a distance" \cite{s}.  But can we say anything quantitative about the nonlocality of quantum correlations?

Let two physicists, Alice and Bob, share many identical pairs of particles with a common origin. Alice measures $a$ or $a^\prime$ on her particles and Bob measures $b$ or $b^\prime$ on his, where $a$, $a^\prime$, $b$ and $b^\prime$ are observables taking values $\pm1$. Their respective measurements on any given pair are spacelike separated.  At some point they pool their data and calculate the correlation functions $C(a,b)$, $C(a,b^\prime)$, $C(a^\prime,b)$ and $C(a^\prime,b^\prime)$.  By definition,
\begin{equation}
C(a,b) = {\rm{p}}_{ab}(1,1) +{\rm{p}}_{ab}(-1,-1) - {\rm{p}}_{ab}(1,-1) -
{\rm{p}}_{ab}(-1,1)~~~,
\label{e2}
\end{equation}
where ${\rm{p}}_{ab}(i,j)$ is the probability that measurements of $a$ and $b$ on a given pair yield $a=i$ and $b=j$.  If the correlations are local, then a certain linear combination of correlations,
\begin{equation}
S_{CHSH} = C(a,b) +C(a,b^\prime) +C(a^\prime,b) - C(a^\prime,b^\prime) ~~~,
\label{BCHSH}
\end{equation}
is bounded by 2 in absolute value \cite{chsh}.  Quantum mechanics, however, obeys---and saturates---``Tsirelson's bound" \cite{ts}, namely $\vert S_{CHSH}\vert \le 2\sqrt{2}$.  The quantum correlations that saturate this bound are $C(a,b) =C(a,b^\prime) =C(a^\prime,b) =\sqrt{2}/2= - C(a^\prime,b^\prime)$, and they are nonlocal.  But it is easy to define correlations that yield $S_{CHSH}=4$:  let $C(a,b) =C(a,b^\prime) =C(a^\prime,b) =1= - C(a^\prime,b^\prime)$.  Since these correlations violate Tsirelson's bound, which is a theorem of quantum mechanics, they are not quantum correlations; we call them ``superquantum" or ``PR-box" correlations.  Now the ideas of Aharonov and Shimony inspire a conjecture:  quantum correlations obey relativistic causality, while superquantum correlations do not.  It sounds like a plausible conjecture; if Alice measures $a$ and obtains $a=1$, quantum correlations do not tell her the result of Bob's measurement on his paired particle; but superquantum correlations tell her that Bob's result was 1 whether he measured $b$ or $b^\prime$.  Apparently, superquantum correlations give {\it too much} information about a spacelike-separated event (Bob's measurement) to be consistent with relativistic causality.  Alas, the conjecture fails, and fails straightforwardly.  Suppose that whether Alice measures $a$ or $a^\prime$, the results $\pm1$ are equally probable (whatever Bob measures); and whether Bob measures $b$ or $b^\prime$, the results $\pm1$ are equally probable (whatever Alice measures). Then Bob and Alice cannot send each other superluminal signals, or even subluminal signals, because Bob's only choice---what to measure, $b$ or $b^\prime$, on his paired particle---does not affect the statistics of Alice's results, and vice versa.  It is straightforward to check that these local probabilities (for Alice and Bob on their own) are compatible with both nonlocal quantum correlations and superquantum correlations, and therefore relativistic causality is compatible with both \cite{PR}.  What a disappointment!  It should not be so easy to disprove such a lovely conjecture!

We might therefore ask, ``If quantum correlations are nonlocal, why aren't they $more$ nonlocal than they are?"  Over the years, others have shown, remarkably, that an additional axiom of communication complexity \cite{o} is sufficient to rule out superquantum correlations, and comes close to ruling out all stronger-than-quantum correlations.  So does a stronger axiom of relativistic causality called ``information causality" \cite{ic}.  However, the physical meaning of communication complexity and information causality is unclear.

But let us take a closer look at superquantum correlations.  We see that if Alice measures $a$ on her particle in a given pair, she knows that Bob, whether he measures $b$ or $b^\prime$ on his particle in the same pair, gets the same result she does; if she measures $a^\prime$ on her particle, she knows that Bob gets the same result she does if he measures $b$ and the opposite result if he measures $b^\prime$.  She can even prepare an ``uncertainty-free" ensemble for Bob, as follows.  Let them share a large number of pairs, and let Alice measure $a$ on all her particles.  If she gets $-1$, she tells Bob to throw out his particle from the same pair; if she gets 1, she tells Bob to keep his particle.  When Alice is done, Bob is left with an ensemble of particles corresponding to $a=1$, and Alice already knows that if Bob measures $b$ on any particle in the ensemble, he will obtain 1, while if he measures $b^\prime$ on any particle in the ensemble, he will obtain 1.  Thus---at least from Alice's point of view---the observables $b$ and $b^\prime$ are not subject to any uncertainty principle on this ensemble.

Now that is strange.  Complementarity---the quantum obligation to choose from among incompatible measurements---is intimately tied to the uncertainty principle \cite{bohr}.  If superquantum correlations are not subject to any uncertainty principle, why can't Bob measure both $b$ $and$ $b^\prime$ on each of his particles?  The inevitable answer is that if Bob could measure both $b$ and $b^\prime$ on even one of his particles, then Alice could certainly send him a superluminal message.  For if Alice chooses to measure $a$, then $b=b^\prime$.  If Alice chooses to measure $a^\prime$, then $b=-b^\prime$.  Thus Bob could read a (one-bit) signal from Alice if he could measure both $b$ and $b^\prime$.  We conclude that, whereas complementarity is intrinsic to quantum mechanics, here in the context of superquantum correlations it is nothing more than a fig leaf---an extraneous item tacked onto the the model to prevent us from seeing Nature's pudenda.

It is not only strange, it is rotten.  No respectable theory should contain such an artificial, cheap fix.  The PR box is rotten.  Unfortunately, by proving it rotten, we have not thereby eliminated it.  What if the universe is rotten?  It is logically possible.

Yet quantum mechanics has a classical limit; in the classical limit, all observables commute.  Let us define macroscopic observables $B$ and $B^\prime$:
\begin{equation}
B={{b_1+ b_2 +\dots+b_N}\over N}~~~~,~~~B^\prime={{b^\prime_1+ b^\prime_2 +\dots+
b^\prime_N}\over N}~~~,
\label{defbbp}
\end{equation}
where $b_m$ and $b^\prime_m$ represent $b$ and $b^\prime$, respectively, on the $m$-th pair. Alice already knows the values of $B$ and $B^\prime$ and, for large $N$, there must be ``weak" measurements \cite{weak} that Bob can make to obtain partial information about both $B$ $and$ $B^\prime$; for, in the classical limit, there can be no complementarity between $B$ and $B^\prime$.  The classical limit is an inherent constraint, a kind of boundary condition, on quantum mechanics, and we should apply this constraint to any generalization of quantum mechanics.  We should not claim that superquantum correlations are more nonlocal than quantum correlations unless they are subject to the same constraints.  Perhaps quantum mechanics would be more nonlocal than it is---even maximally nonlocal---if it were not constrained to have a classical limit.  In this sense, the axiom of a classical limit is minimal: if it applies to quantum correlations, it should apply also to proposed generalizations of quantum correlations, including superquantum correlations.  Thus superquantum correlations, too, must have a classical limit.

We therefore assume that for large enough $N$, there must be some measurements Bob can make to obtain partial information about $both$ $B$ $and$ $B^\prime$.  And now the game changes. It is true that $a=1$ and $a=-1$ are equally likely, and so the average values of $B$ and $B^\prime$ vanish, whether Alice measures $a$ or $a^\prime$.  But if she measures $a$ on each pair, then typical values of $B$ and $B^\prime$ will be $\pm 1/\sqrt{N}$ (but possibly as large as $\pm 1$) and correlated.  If she measures $a^\prime$ on each pair, then typical values of $B$ and $B^\prime$ will be $\pm 1/\sqrt{N}$ (but possibly as large as $\pm 1$) and $anti$-correlated.  Thus Alice can signal to Bob by consistently choosing whether to measure $a$ or $a^\prime$.  This claim is delicate because the large-$N$ limit in which $B$ and $B^\prime$ commute is also the limit that suppresses the fluctuations of $B$ and $B^\prime$.  To insure that Bob has a good chance of measuring $B$ and $B^\prime$ accurately enough to determine whether they are correlated or anti-correlated, $N$ may have to be large and therefore the fluctuations in $B$ and $B^\prime$ will be small.  However, Alice and Bob can repeat this experiment (on $N$ pairs at a time) as many times as it takes to give Bob a good chance of catching and measuring large enough fluctuations.  They can repeat the experiment exponentially many times.  Alice and Bob's expenses and exertions are not our concern.  Relativistic causality does not forbid superluminal signalling only when it is cheap and reliable.  Relativistic causality forbids superluminal signalling altogether.

It might be that the errors and uncertainties in Bob's measurements are so large that some, or even most, of the (anti-)correlations between $B$ and $B^\prime$ that he detects are erroneous.  We cannot specify exactly how the approach to the classical limit depends on $N$.  But this is no objection.  What matters is only that when Bob detects a correlation, it is more likely that Alice measured $a$ than when he detects an anti-correlation.  If it were not more likely, it would mean that Bob's measurements yield zero information about $B$ or about $B^\prime$, contradicting the fact that there is a classical limit in which $B$ and $B^\prime$ are jointly measurable.

As a concrete example, let us suppose Bob considers only those sets of $N$ pairs in which $B=\pm1$ and $B^\prime=\pm 1$.  The probability of obtaining $B=1$ is $2^{-N}$.  But if Alice is measuring $a$ consistently, the probability of obtaining $B=1$ $and$ $B^\prime =1$ is also $2^{-N}$, and not $2^{-2N}$, while the probability of obtaining $B=1$ and $B^\prime =-1$ vanishes.  If Alice is measuring $a^\prime$ consistently, the probabilities are reversed.  Thus with unlimited resources, Alice can send a (superluminal) signal to Bob. Superquantum (PR-box) correlations are $not$ consistent with relativistic causality in the classical limit.

We have ruled out superquantum correlations.  To recover quantum correlations, however, we have to rule out all stronger-than-quantum correlations, i.e. we have to derive Tsirelson's bound from the three axioms of nonlocality, relativistic causality, and the existence of a classical limit.  The proof \cite{max} is technical, but here are the main points.

Consider sets of $N$ pairs exhibiting superquantum correlations.  No matter what Alice measures, the averages $\langle B\rangle$, $\langle B^\prime \rangle$ and therefore also $\langle B+B^\prime\rangle$ tend towards 0.  But if Alice measures $a^\prime$ on all the pairs, then $B+B^\prime =0$ identically for each set of $N$ pairs, since $a^\prime$ and $b$ are perfectly correlated while $a^\prime$ and $b^\prime$ are perfectly anti-correlated.  If Alice measures $a$ on all the pairs, then the values of $B+B^\prime$ on successive sets of $N$ pairs fall in a binomial distribution centered at 0.  As long as Bob has $some$ information about $B+B^\prime$ and its variance, he will ultimately be able to read Alice's one bit of information, i.e. whether Alice measures $a$ or $a^\prime$ on all her pairs.  Now suppose that the (anti-)correlations are not perfect but near-perfect.  Then the variance of $B+ B^\prime$ will not vanish if Alice measures $a^\prime$ on all her pairs, but will still be significantly smaller than if Alice measures $a$ on all her pairs, and Bob will still ultimately be able to read Alice's one bit of information.  As the (anti-)correlations get weaker, however, the variances of $B+ B^\prime$ and $B- B^\prime$ will approach each other, until, at some critical correlation, Bob will not be able to read Alice's one bit.  A natural guess is that this critical correlation corresponds to the maximum quantum correlation $\sqrt{2}/2$, the correlation that saturates Tsirelson's bound.  Surprisingly, this guess fails:  the critical correlation is 1/2, the correlation that saturates the CHSH inequality!  How can it be?

We arrive at a paradox:  quantum correlations conform to relativistic causality, yet the calculation of the maximal correlation consistent with relativistic causality passes right by quantum correlations and arrives at local correlations!

The root of this paradox is that the PR box, and more generally any nonlocal box, yields $\langle B\rangle$ and $\langle B^\prime \rangle$ and therefore also $\langle B\pm B^\prime\rangle$, and yields the variances of $B$ and $B^\prime$; but it does not yield the variances of $B\pm B^\prime$ unless we tacitly assume that the $b_m$ and $b_m^\prime$ combine in a certain way.  When we calculate the variances of $B\pm B^\prime$ $without$ this assumption---that is, when we think ``outside the box"---we do indeed obtain Tsirelson's bound.  Indeed, we obtain more than Tsirelson's bound.  If we examine the tacit assumption that led to the paradox, we see that it forces $b_m$ and $b_m^\prime$ to add as $scalars$.  If $b_m$ and $b_m^\prime$ are allowed to add as vectors, they $can$ saturate Tsirelson's bound.  So this derivation of Tsirelson's bound shows that Hilbert space is implicit in quantum correlations.

\begin{acknowledgments}

For over two decades I have had the great good fortune to work with Professor Yakir Aharonov, learning from his penetrating questions, his mastery of quantum and statistical fluctuations, his subtle formulations such as weak measurement and weak values, and his countless other insights.

\end{acknowledgments}


\begin{references}

\bibitem{QP} Y. Aharonov and D. Rohrlich, {\it Quantum Paradoxes: Quantum Theory for the
Perplexed} (Weinheim: Wiley-VCH), 2005.

\bibitem{hp} H. Poincar\'e, {\it Sechs Vortr\"age aus der Reinen Mathematik und
Mathematischen Physik} (Leipzig:  Teubner), 1910, trans. and cited in A. Pais, {\it `Subtle is the Lord...':  the Science and Life of Albert Einstein} (New York:  Oxford University Press), 1982, pp. 167-8.

\bibitem{AB}Y. Aharonov and D. Bohm, {\it Phys. Rev.} {\bf 115} (1959) 485.

\bibitem{bell} J. S. Bell, {\it Physics} {\bf 1}, 195 (1964).

\bibitem{chsh} J. F. Clauser, M. A. Horne, A. Shimony, and R. A. Holt, {\it Phys. Rev.
Lett.} {\bf 23}, 880 (1969).

\bibitem{nleq}Y. Aharonov, H. Pendleton and A. Petersen, {\it Int. J. of Theor. Phys.} {\bf
2}, 213 (1969); Y. Aharonov, in {\it Proc. of the Int. Symp. on the Foundations of Quantum
Mechanics}, Tokyo, 1983, p. 10.  See also Y. Aharonov and D. Rohrlich, {\it op. cit.}, Chaps. 5, 6 and 13.

\bibitem{s}A. Shimony, in {\it Foundations of Quantum Mechanics in Light of the New
Technology}, S. Kamefuchi et al. eds. (Tokyo:  Japan Physical Society), 1983, p. 225; A.
Shimony, in {\it Quantum Concepts of Space and Time}, R. Penrose and C. Isham, eds. (Oxford: Clarendon Press), 1986, p. 182.

\bibitem{ts}B. S. Tsirelson (Cirel'son), {\it Lett. Math. Phys.} {\bf 4}, 93 (1980).

\bibitem{PR} S. Popescu and D. Rohrlich, {\it Found. Phys.} {\bf 24}, 379 (1994).  See also
D. Rohrlich, in (The Frontiers Collection) {\it Probability in Physics}, eds. Y. Ben-Menahem and M. Hemmo (Berlin: Springer), 2012, pp. 187-200.

\bibitem{o} W. van Dam, {\it  Nonlocality \& Communication Complexity} (Ph.D. thesis),
Oxford University (2000); preprint quant-ph/0501159 (2005); D. Dieks, {\it Phys. Rev.} {\bf A66}, 062104 (2002); H. Buhrman and S. Massar, {\it Phys. Rev.} {\bf A72}, 052103 (2005);  J. Barrett and S. Pironio, {\it Phys. Rev. Lett.} {\bf 95}, 140401 (2005); G. Brassard, H. Buhrman, N. Linden, A. A. M{\'e}thot, A. Tapp and F. Unger, {\it Phys. Rev. Lett.} {\bf 96}, 250401 (2006); J. Barrett, {\it Phys. Rev.} {\bf A75}, 032304 (2007); D. Gross, M. M{\"u}ller, R. Colbeck and O. C. O. Dahlsten, {\it Phys. Rev. Lett.} {\bf 104}, 080402 (2010).

\bibitem{ic}M. Paw{\l}owski et al., {\it Nature} {\bf 461}, 1101 (2009).

\bibitem{bohr}N. Bohr, in {\it Albert Einstein: Philosopher--Scientist}, ed. Paul A.
Schilpp (New York:  Tudor Pub. Co.), 1951, pp. 201-41.

\bibitem{weak}Y. Aharonov, D. Z. Albert, and L. Vaidman, {\it Phys. Rev. Lett.} {\bf 60},
1351 (1988); see also Y. Aharonov and D. Rohrlich, {\it op. cit.}, Chaps. 16-17.

\bibitem{max}D. Rohrlich, ``Stronger-than-quantum bipartite correlations violate
relativistic causality in the classical limit", submitted to {\it Phys. Rev. Lett.}

\end{references}
\end{document}